\documentclass[12pt]{article}

\usepackage{graphicx,euscript}

\def \bea{\begin{eqnarray}}
\def \beq{\begin{equation}}
\def \bo{B^0}
\def \b{{\cal B}}

\def \eea{\end{eqnarray}}
\def \eeq{\end{equation}}

\def \lrr{\lambda^L_{\rho \rho}}

\def \ob{\overline{B}^0}

\newcommand*{\R}{\ensuremath{\mathcal{R}}}

\begin{document}

\renewcommand{\thefootnote}{\fnsymbol{footnote}}
\setcounter{footnote}{1}

\begin{titlepage}

\begin{flushright}
PITHA 06/05\\
TECHNION-PH-2006-05\\
hep-ph/0604005 \\
2 April 2006
\end{flushright}

\vspace{1cm}
\begin{center}

{\large\bf\boldmath 
A precise determination of $\alpha$ using $B^0\to\rho^+\rho^-$ 
and $B^+\to K^{*0}\rho^+$}\footnote{To be published in Physics 
Letters B.}

\vspace{0.5cm}

M.~Beneke${}^{\,(a)}$, 
M.~Gronau${}^{\,(b)}$, 
J. Rohrer${}^{\,(a)}$, 
M. Spranger${}^{\,(b)}$\\[0.3cm]
{\sl ${}^{(a)\,}$Institut f\"ur Theoretische Physik E, RWTH Aachen,\\ 
D -- 52056 Aachen, Germany\\
${}^{(b)\,}$Department of Physics, Technion-Israel Institute of Technology, \\
Technion City, Haifa 32000, Israel}
\end{center}

\vspace{0.8cm}
\begin{abstract}
\vspace{0.2cm}
\noindent 
The effect of the penguin amplitude on
  extracting $\alpha$ from CP asymmetries in $B^0\to \rho^+\rho^-$ decays 
  is studied using information on the SU(3)-related penguin amplitude in 
  $B^+\to K^{*0}\rho^+$. Conservative bounds on non-factorizable SU(3) 
  breaking, small amplitudes, and the strong phase difference between 
tree and penguin amplitudes, are shown to reduce the error in 
  $\alpha$ in comparison with the one obtained using isospin symmetry in 
  $B\to\rho\rho$. Current measurements imply $\alpha = 
  [90 \pm 7~({\rm exp})^{+2}_{-5}~({\rm th})]^\circ$.
\end{abstract}

\end{titlepage}

\renewcommand{\thefootnote}{\arabic{footnote}}
\setcounter{footnote}{0}


{\bf 1.} 
A major purpose for studying $B$ and $B_s$ decays is
achieving great precision in Cabibbo-Kobayashi-Maskawa (CKM)
parameters and providing precision tests for the Kobayashi-Maskawa
mechanism of CP violation~\cite{Kobayashi:1973fv}.  CP asymmetries in
properly chosen decays can be related with high precision to the
angles $\beta, \gamma$ and $\alpha=\pi-\beta-\gamma$ of the CKM
unitarity triangle~\cite{Gronau:2005cz,Charles:2004jd,Bona:2005vz}.

Currently the most precise single (hadronic-theory independent)
determination of $\alpha$ or $\gamma$ is based on the 
CP asymmetries $C_L$ and $S_L$ in $B^0\to\rho^+\rho^-$ and on isospin symmetry 
in the $B\to\rho\rho$ system. The observation that  
the $\rho$ mesons in $B^0\to\rho^+\rho^-$ are nearly entirely longitudinally
polarized~\cite{Zhang:2003up,Aubert:2003mm,Aubert:2005nj,Somov:2006sg}
has simplified this study to becoming equivalent to an isospin analysis in 
$B\to\pi\pi$~\cite{Gronau:1990ka}. Using isospin triangle
inequalities~\cite{Grossman:1997jr,Gronau:2001ff} the current upper limit
$\b(B\to\rho^0\rho^0)< 1.1\times 10^{-6}$~\cite{Aubert:2004wb} 
implies $\alpha=(96\pm 13)^\circ$ \cite{Aubert:2005nj,Somov:2006sg}, which
includes an intrinsic error of $11^\circ$ from the penguin 
amplitude alone, and only $7^\circ$ originating in the measured CP 
asymmetries $C_L$ and $S_L$.

The purpose of this Letter is to suggest an alternative way for
studying the penguin amplitude effect on measuring $\alpha$ in
longitudinally polarized $B^0\to\rho^+\rho^-$. We relate the penguin
amplitude in this process to the longitudinal amplitude in $B^+\to
K^{*0}\rho^+$ which is dominated by a $\Delta S=1$ penguin
contribution.  The resulting error in $\alpha$ is shown to be smaller
than in the isospin analysis of the three $B\to
\rho\rho$ decays. This will be argued to be the case in spite of a 
larger theoretical uncertainty caused by flavour SU(3) and further 
approximations entering the determination of the penguin amplitude in 
$B^0\to\rho^+\rho^-$. The main point is
that a large relative uncertainty in the penguin amplitude leads
to only a small uncertainty in $\alpha$, once the penguin amplitude is
established to be small.
Applications of flavour SU(3) to $B^0\to\pi^+\pi^-$
and $B^0\to K^+\pi^-$ or $B^+\to K^0\pi^+$~\cite{Gronau:2004ej,Buras:2004ub}  
involve a somewhat larger theoretical uncertainty in $\alpha$ 
because the ratio of penguin to tree amplitudes is considerably larger 
in $B^0\to\pi^+\pi^-$ than in $B^0\to\rho^+\rho^-$.

{\bf 2.} The amplitude for longitudinally polarized $\rho$ mesons can
generally be written as
\begin{equation}\label{Arhorho}
  A_L(B^0\to\rho^+\rho^-) = Te^{i\gamma} + P e^{i\delta}~.
\end{equation}
By convention $T$ and $P$ are positive, involving the magnitudes of the
CKM factors $V^*_{ub}V_{ud}$ and $V^*_{cb}V_{cd}$, and 
the strong phase $\delta$ lies in the range $-\pi < \delta \le \pi$.
Time-dependence
for longitudinal polarization is described in terms of two CP
asymmetries $C_L$ and $S_L$~\cite{Gronau:1989ia},
\begin{equation}
  \frac {\Gamma_L({\bar B}^0(t) \to \rho^+\rho^-) - 
         \Gamma_L(\bo(t)\to\rho^+\rho^-)}
        {\Gamma_L({\bar B}^0(t) \to \rho^+\rho^-) +
         \Gamma_L(\bo(t)\to\rho^+\rho^-)}
    = -C_L\cos(\Delta mt) + S_L\sin(\Delta mt)~.
\end{equation}
The asymmetries $C_L$ and $S_L$ are given by
\begin{equation} \label{eqn:CSpipi}
  C_L = \frac{1 - |\lrr|^2}{1 + |\lrr|^2}~, \qquad
  S_L = \frac{2 {\rm Im}(\lrr)}{1 + |\lrr| ^2}~,
\end{equation}
where
\begin{equation}
  \lrr \equiv 
     e^{-2i \beta} 
     \frac{A_L({\bar B}^0 \to \rho^+ \rho^-)}%
          {A_L(B^0 \to \rho^+ \rho^-)}~.
\end{equation}
Substituting (\ref{Arhorho}) into these definitions, one obtains
\begin{eqnarray}
  \label{C}
  C_L & = & \frac{2r\sin\delta\sin(\beta +\alpha)}%
                 { 1 - 2r\cos\delta\cos(\beta + \alpha) + r^2}~, \\
  \label{S}
  S_L & = & \frac{\sin 2\alpha + 2r\cos\delta\sin(\beta-\alpha) - 
                  r^2\sin 2\beta}%
                 { 1 - 2r\cos\delta\cos(\beta + \alpha) + r^2}~,
\end{eqnarray}
where
\begin{equation}
  r \equiv \frac{P}{T} > 0
\end{equation}
is the ratio of the penguin to the tree amplitude.

In the absence of a penguin amplitude ($r=0$) one has
$C_L=0,~S_L=\sin 2\alpha$.  For small values of $r$ one finds
\begin{eqnarray}\label{leadingC}
  C_L & = & 2r\sin\delta\sin(\beta + \alpha) + {\cal O}(r^2)~, \\
  \label{leadingS}
  S_L & = & \sin 2\alpha + 2r\cos\delta\sin(\beta + \alpha)\cos 2\alpha
            + {\cal O}(r^2)~.
\end{eqnarray}
Given the value of $\beta$~\cite{Charles:2004jd}, 
\begin{equation}\label{beta}
  \beta = (21.7^{+1.3}_{-1.2})^\circ~,
\end{equation}
the two measurables $C_L$ and $S_L$ provide two equations for the weak
phase $\alpha$ and for the two hadronic parameters $r$ and $\delta$.
An additional constraint on $r, \delta$ and $\alpha$ is needed in
order to determine the weak phase.

We will use the decay rate for a longitudinally polarized state in
$B^+\to K^{*0}\rho^+$. The magnitude of the penguin amplitude
dominating this process is related by flavour SU(3) to the magnitude of
the penguin amplitude in $B^0\to \rho^+\rho^-$~\cite{Gronau:1994rj}.
An additional constraint may, in principle, be obtained using the
process $B^0\to K^{*+}\rho^-$ for longitudinally polarized final
states. In this case SU(3) relations apply to $P, T$ and $\delta$ and
their SU(3) counterparts in $B^0\to K^{*+}\rho^-$. However, so far
only an upper limit has been measured for the decay rate of this
process~\cite{Aubert:2004dp}, and further information about the
longitudinal fraction would be required.

The amplitude squared for decays into longitudinally polarized
$K^{*0}\rho^+$ final states can be written as
\begin{equation}\label{K*rho}
  |A_L(B^+\to K^{*0}\rho^+)|^2_{\rm CP-av.} = 
    \left(\frac{|V_{cs}|}{|V_{cd}|}\frac{f_{K^*}}{f_\rho}\right)^2
    F P^{\,2} = 21.4 \,F P^{\,2}~,
\end{equation}
where $f_{\rho}=(209\pm 1)\,$MeV and $f_{K^*}=(218\pm 4)\,$ MeV are the
vector meson decay constants~\cite{Beneke:2003zv}, and $P$ is the 
penguin amplitude defined in (\ref{Arhorho}). 
This equation \emph{defines} a parameter $F$, 
which equals one when neglecting non-factorizable SU(3) breaking
corrections (i.e.~SU(3) breaking not in decay constants and form 
factors) in magnitudes of penguin amplitudes, and other contributions
as discussed below. We now define a
ratio of CP-averaged decay rates,
\begin{equation}
  \R \equiv 
    \left(\frac{|V_{cd}|}{|V_{cs}|}\frac{f_\rho}{f_{K^*}}\right)^{\!2}
    \,\frac{\Gamma_L(B^+\to K^{*0}\rho^+)+
          \Gamma_L(B^-\to \bar K^{*0}\rho^-)}%
         {\Gamma_L(\bo\to\rho^+\rho^-)+
          \Gamma_L(\ob\to \rho^+\rho^-)}~,
\end{equation}
whose measurement provides a third constraint on $r, \delta$ and $\alpha$:
\begin{equation}\label{calR}
  \R = \frac{F r^2}{1 - 2r\cos\delta\cos(\beta + \alpha) + r^2}~.
\end{equation}
Eqs.~(\ref{C}), (\ref{S}) and (\ref{calR}) give the three observables
$C_L$, $S_L$ and $\R$ in terms of $r$, $\delta$ and $\alpha$.
Assuming $F$ is known permits a solution for $\alpha$ up to discrete 
ambiguities.

{\bf 3.} We proceed to discuss the parameter $F$ which, crudely
speaking, relates the penguin amplitude squared in $B\to K^*\rho$ to the
one in $B\to\rho\rho$. At the amplitude level, the parameter $F$
involves several effects. In addition to non-factorizable
SU(3)-breaking it includes corrections from a color-suppressed
electroweak penguin amplitude, penguin annihilation
contributions~\cite{Gronau:1994rj,Beneke:2003zv}, and a doubly
CKM-suppressed penguin amplitude. These corrections are usually
thought to be small, so that $F$ is expected to be near unity.  We
shall discuss each of the four corrections in turn.

The neglect of non-factorizable SU(3)-breaking corrections is implicit
in all applications of SU(3) flavour symmetry to hadronic $B$ decays.
Within the current experimental uncertainties there is no evidence for
the need of such a correction in the analysis of $B$ decays to final
states involving two pseudoscalar mesons ($B\to
PP$)~\cite{Gronau:2004ej,Buras:2004ub,Chiang:2004nm}  
and decays into a pair of pseudoscalar and vector mesons  
($B\to VP$)~\cite{Chiang:2003pm}. 
We assume that final states with two vector mesons ($B\to VV$) 
are no different in this respect. In the QCD factorization approach
\cite{Beneke:1999br,Beneke:2000ry}, non-factorizable SU(3) breaking
corrections arise primarily from differences in light-cone
distribution amplitudes of $K^*$ and $\rho$. This correction is
unlikely to exceed $15\%$ at the amplitude level.  The doubly
CKM-suppressed penguin amplitude proportional to
$|V^*_{ub}V_{us}|/|V^*_{cb}V_{cs}| \sim 0.02$ is negligible, since no
plausible dynamical mechanism is known which would enhance  this
amplitude without enhancing the dominant penguin amplitude.

More important are the colour-suppressed electroweak penguin 
amplitudes in both $B^0\to \rho^+\rho^-$ and $B^+\to K^{*0}\rho^+$, 
usually denoted $P^c_{EW}$~\cite{Gronau:1994rj} 
or $\alpha_{4,\rm EW}^{c}$~\cite{Beneke:2003zv}, and a penguin-annihilation
amplitude in $B^0\to \rho^+\rho^-$, denoted 
$PA$~\cite{Gronau:1994rj} or $2\beta_4^c$~\cite{Beneke:2003zv}.
Since the dominant QCD penguin amplitude is smaller for $B\to VV$ than
for $B\to PP$, these two contributions are 
comparatively more significant in $B\to VV$ than in $B\to PP$ decays 
where they are 
often neglected. For orientation, a QCD factorization calculation of 
$B\to VV$ decays~\cite{Beneke:2005we} gives that the
colour-suppressed electroweak penguin correction decreases $F$ by
about 0.1. The penguin-annihilation effect is about $-0.3$, and thus 
turns out to be the largest contributor to 
$F-1$ in spite of being formally suppressed by
$1/m_b$~\cite{Beneke:2000ry}. 
A global SU(3) fit to all $B\to VV$ decays, which requires more data, 
may eventually be able to check the size of penguin annihilation 
amplitude in these decays. One consequence of this 
contribution~\cite{Gronau:1994rj} is a non-negligible branching ratio 
for longitudinally polarized $B_s \to \rho^+\rho^-$ decays, 
on the order of a few times $10^{-7}$ \cite{rohrer}.

A random scan through the input parameter space in the QCD
factorization calculation \cite{Beneke:2003zv,Beneke:2005we} that 
includes all four effects yields a nearly
Gaussian distribution for $F$ with $F=0.65\pm 0.36$. This
estimate depends crucially on whether the annihilation model adopted
in \cite{Beneke:2000ry} predicts correctly the magnitude and sign of $PA$ or 
$2\beta_4^c$. Since we would not like to rely on this assumption, we
shall adopt the wider range,
\begin{equation}
  \label{F}
  0.3 \le F \le 1.5~. 
\end{equation}
Thus we are allowing a variation in $P^2$ by a factor of five and in
$P$ by a factor larger than two.
We will study below the sensitivity of the extracted error in $\alpha$
to this rather conservative range, showing that in spite of the 
large theoretical uncertainty allowed in $F$ the determination of 
$\alpha$ is quite precise since data 
requires $r$ to be small.

{\bf 4.} 
We now describe the experimental status of $C_L$,
$S_L$ and $\R$. The most recent measurements of $C_L$ and $S_L$ by the
BABAR~\cite{Aubert:2005nj} and BELLE~\cite{Somov:2006sg} collaborations
are
\begin{eqnarray}
  C_L  & = & \left\{ 
    \begin{array}{cl}
     &-0.03 \pm 0.18 \pm 0.09~, \\
     & \phantom{-}0.00 \pm 0.30 \pm 0.09~. 
    \end{array} \right.\\
  S_L  & = & \left\{
    \begin{array}{cl}
     &  -0.33 \pm 0.24^{+0.08}_{-0.14}~, \\
     &  \phantom{-}0.08 \pm 0.41 \pm 0.09~.      
    \end{array} \right.
\end{eqnarray}
Here (and below) BABAR and BELLE values are represented 
by upper and lower entries, respectively.
These values imply the averages~\cite{Group:2006bi},
\begin{equation}\label{CSexp}
  C_L = -0.03 \pm 0.17~,\qquad
  S_L = -0.21 \pm 0.22~.
\end{equation}

In order to compute $\R$ we use the CP-averaged branching ratios 
(given in units of
$10^{-6}$) and longitudinal polarization fractions, as obtained by 
BABAR~\cite{Aubert:2005nj,Aubert:2004zr,Aubert:2004qb} and 
BELLE~\cite{Somov:2006sg,Zhang:2005iz},
\begin{eqnarray}
  \b(\rho^+\rho^-) &=& \left\{
    \begin{array}{cl}
      30 \pm 4 \pm 5~, \\
      22.8 \pm 3.8^{+2.3}_{-2.6}~,
    \end{array} \right.
  \hskip1.2cm
  f_L(\rho^+\rho^-) = \left\{
    \begin{array}{cl}
      0.978\pm 0.014^{+0.021}_{-0.029}~,\\
      0.941^{+0.034}_{-0.040}\pm 0.030~,
    \end{array} \right.
  \label{Brs}\\[0.3cm]
  \b(K^{*0}\rho^+) &=& \left\{
    \begin{array}{cl}
      17.0 \pm 2.9 \pm 2.0 ^{+0.0}_{-1.9}~,\\
       8.9 \pm 1.7 \pm 1.2~,
    \end{array} \right.
  \hskip0.2cm
  f_L(K^{*0}\rho^+) = \left\{
    \begin{array}{cl}
      0.79 \pm 0.08 \pm 0.04\pm 0.02~,\quad\\
      0.43\pm 0.11^{+0.05}_{-0.02}~.
    \end{array}\right.
  \label{fLs}
\end{eqnarray}
Using the $B^+/B^0$ lifetime ratio
$\tau_+/\tau_0=1.076\pm 0.008$~\cite{Group:2006bi}, this implies 
\begin{equation}
  \R = \left\{
    \begin{array}{cl}
      0.0199 \pm 0.0065~,\\
      0.0077 \pm 0.0032~.
    \end{array} 
    \label{Rvalues}
\right.
\end{equation}
The two values, representing BABAR and BELLE results, are not in good 
agreement with each other. The difference of $1.7 \sigma$ originates mainly 
from a difference by a factor 3.5 between the two measurements of longitudinal 
$B^+\to K^{*0}\rho^+$ branching ratios. The weighted average of the two values
in (\ref{Rvalues}) is ${\cal R}= 0.0101 \pm 0.0029$. 
Calculating $\R$ from the averages 
of (\ref{Brs}) and (\ref{fLs}), we find a slightly larger value 
(implying a slightly larger error in the extracted value of $\alpha$),
\begin{equation}\label{Rexp}
  \R = 0.0125 \pm 0.0031~.
\end{equation}
We will use this value, the error of which does not include a scaling 
factor to account for the disagreement between the BABAR and BELLE 
measurements in (\ref{fLs}). We may expect this disagreement 
to disappear in the future. Note, however, that 
the effect of the experimental error in (\ref{Rexp}) on the extracted 
value of $\alpha$ is smaller than that of the theoretical uncertainty given by 
the wide range (\ref{F}) for $F$ to which ${\cal R}$ is proportional 
[see~(\ref{calR})].

{\bf 5.} 
For given values of $C_L$, $S_L$, $\R$ and  fixed $F$, 
(\ref{C}),~(\ref{S}) and (\ref{calR}) can be solved numerically. 
The solutions exhibit an eightfold ambiguity for $\delta$ and $\alpha$
in the range $-\pi < \delta \le \pi, -\pi \le \alpha \le \pi$, which 
can be understood and resolved into three independent invariance 
transformations ($2^3 =8$) obeyed by (\ref{C}),~(\ref{S}) and
(\ref{calR}):
\begin{eqnarray}
\mbox{(i)} && \delta \to \pi +\delta,~\alpha \to \pi + \alpha,~r \to r~,
\nonumber\\[0.2cm]
\mbox{(ii)} && \delta \to \pi -\delta,~\alpha \to 
 \alpha\,[1 + {\cal O}(r)],~r \to r\,[1+{\cal O}(r)]~,
\\[0.2cm]
\mbox{(iii)} && \sin\delta \to -\sin\delta\,
\frac{\sin(\beta+\alpha)}{\cos(\beta-\alpha)}
\,[1+{\cal O}(r)],~\alpha \to (3\pi/2 - \alpha)\,[1+{\cal O}(r)],
\nonumber\\
\mbox{} &&r \to r\,[1+{\cal O}(r)]~.
\nonumber
\end{eqnarray}
The first transformation is an exact symmetry of the three equations,
leading to unphysical values of $\alpha$ larger than $\pi$ or negative.
These four solutions can be discarded, leaving four solutions in
the range $0 \le \alpha \le \pi$. 
The second and third transformations, $\delta\to \pi -\delta$ and 
$\alpha \to 3\pi/2 -\alpha$,  do not change the leading terms in $r$ 
for $C_L$, $S_L$ and ${\cal R}$ given in ({\ref{leadingC}), 
(\ref{leadingS}) and (\ref{calR}). 
Including the non-leading terms in the expressions for $C_L,~S_L$ and 
${\cal R}$ implies a correction of order $r$ in $r$, 
and corresponding corrections in $\alpha$ and $\delta$, 
in the transformations (ii) and (iii).

\begin{table}
  \begin{center}
    \begin{tabular}{cccc}
    solution &
      $\alpha\,[{}^\circ]$ & 
      $r$ & 
      $\delta\,[{}^\circ]$ (central)
      \\
      \hline\hline
      &&& \\[-0.3cm]
      (1) &
      $\phantom{1}89.8^{ +7.2}_{-6.7}$ & 
      $0.124^{+0.015}_{-0.017}$ &
      $\phantom{7}-8$
      \\
      &&& \\[-0.4cm]
      (2) &
      $101.6^{ +6.3}_{-6.1}$ &
      $0.111^{+0.012}_{-0.014}$ &
      $-172$ 
      \\
      &&& \\[-0.4cm]
      \hline
      &&& \\[-0.3cm]
     (3) &
      $175.8^{ +7.3}_{-9.4}$ &
      $0.131^{+0.019}_{-0.038}$ &
      $\phantom{-1}28$
      \\
      &&& \\[-0.4cm]
      (4) &
      $\,\,172.5^{+10.6}_{-6.1}$ &
      $0.107^{+0.043}_{-0.014}$ &
      $\phantom{-}152$
      \\
    \end{tabular}
  \end{center}
  \caption{Four solutions for $\alpha,~r$ and $\delta$ 
      corresponding to $F=0.9$. \label{tab:eightsolutions}}
\end{table}

Keeping the theoretical parameter $F$ fixed at its central value, 
$F=0.9$, and using the measurements given
in~(\ref{CSexp}) and (\ref{Rexp}) for $C_L, S_L$ and ${\cal R}$, we
solve for $r, \delta$, and $\alpha$ in the physical range 
$0 \le \alpha \le \pi$. The four solutions obtained within $\chi^2=1$ 
contours for $(C_L,S_L,\R)$ are given in Table I. {\em An important 
observation is the small value of $r$, in the range
$0.10-0.13$, which is implied by the small measured value 
of ${\cal R}$.} While the errors obtained for $\alpha$ and $r$ are 
reasonably small, we only quote central values for $\delta$ for 
which the errors are large. (See Figure 2 and discussion
below.) We see that, as implied by the transformation (ii), the
solutions (1,3) transform to the solutions (2,4) under 
$\delta \to \pi - \delta$. The change in $\alpha$ under this 
transformation is first order in $r$ and is therefore rather small. 
A much larger change in $\alpha$ is implied by the transformation 
(iii), $\alpha \to 3\pi/2 -\alpha$, 
replacing $(1) \to (3), (2) \to (4)$. Solutions (3) and (4) are 
excluded by the measured value
of $\beta$ in (\ref{beta}) and by $\alpha +\beta +\gamma =\pi$. 

The two  remaining solutions, (1) and (2), both lying in the vicinity 
of $\alpha = \pi/2$, can be distinguished by their values of the strong 
phase $\delta$. It is clear from (\ref{leadingS}),
where $\sin(\beta +\alpha)\cos 2\alpha <0$ holds for both solutions, 
that the smaller and larger solutions for $\alpha$ correspond to 
$\cos\delta>0$ and $\cos\delta < 0$, respectively. In the QCD factorization 
approach~\cite{Beneke:1999br,Beneke:2000ry} 
the phase $\delta$ is predicted to be small, being suppressed by
$1/m_b$ or $\alpha_s(m_b)$. This excludes solution (2) leaving as the  
single solution the value $\alpha = 89.8^{ +7.1}_{-6.7}$. 
Note that we do not need to assume that the phase $\delta$ is small. 
It is sufficient to exclude values of $\delta$ near $\pm 180^\circ$.
(A more precise requirement, depending on experimental errors on $\delta$,  
will be given when discussing Figure 2 below.)
The error $\pm 7^\circ$ in $\alpha$ is essentially the 
same as the error obtained in $\alpha_{\rm eff}$ using the isospin 
method~\cite{Aubert:2005nj}. This is not surprising, since by fixing 
the value of the parameter $F$ to the central value in the range 
(\ref{F}) we have restricted the origin of the error in the extracted 
value of $\alpha$ to experimental errors in the asymmetries $C_L$ and 
$S_L$. The effect of the error in ${\cal R}$ given by (\ref{Rexp}) is 
relatively minor. 

\begin{figure}[t]
  \centering
  \includegraphics[width=10cm]{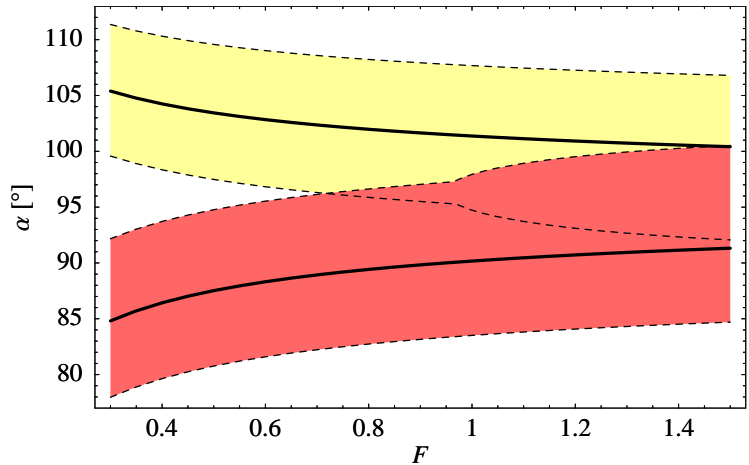}
  \caption{Dependence of $\alpha$ on $F$ for the two solutions in Table I:
(1) lower band, (2) upper band. The two bands denote experimental errors 
from $C_L,~S_L$ and ${\cal R}$.
  \label{fig:plot}  }
\end{figure}

The only theoretical error in our method (up to the discrete phase 
choice) originates in the parameter $F$. 
We now discuss the extraction of $\alpha$ for the entire range of
$F$ given in (\ref{F}), focusing on the two solutions (1) and (2) 
near $\alpha = \pi/2$. In Figure~\ref{fig:plot} we show the dependence
of these solutions on the parameter $F$.
The lower and upper solid dark lines, corresponding to solutions (1) and (2)
respectively,  use central values for $C_L,~S_L$ and ${\cal R}$. 
The bands around these two lines give experimental errors originating
in these three measurements. Focusing on the theoretical error from
$F$ alone, we consider values of $\alpha$ along the 
dark solid lines, 
comparing values at $F=0.9$ with values at $F=0.3$ and
$F=1.5$. We find the variation in the lower and upper solutions to be 
given by $(89.8^{+1.5}_{-5.0})^\circ$ and
$(101.6^{+3.7}_{-1.2})^\circ$, respectively.
We discard again the second solution on the basis of involving values 
of $\delta$ in the neighborhood of $\pi$ rather than near zero. 
Including the experimental error from 
Table I and the above theoretical uncertainty from $F$, we conclude
\beq\label{alpha}
\alpha = [89.8^{+7.2}_{-6.7}~({\rm exp})^{+1.5}_{-5.0}~({\rm th})]^\circ~.
\eeq
We note that {\em the theoretical error, following from the range
  (\ref{F}) in $F$ and the preference for one of the two theoretically
  possible solutions, is considerably smaller than the error of
  $11^\circ$ in $\alpha$ obtained from an upper bound on 
$|\alpha-\alpha_{\rm eff}|$ by applying the isospin triangle analysis 
to $B\to\rho\rho$~\cite{Aubert:2005nj}.} It is worth recapitulating 
the origin of this small error: Data on ${\cal R}$ implies that 
the penguin correction is small. Once this is established 
the relation $S_L=\sin 2\alpha$ receives only small corrections, 
and since $\sin 2\alpha$ is rapidly varying near $\alpha=\pi/2$ 
even a significant error in $S_L$ translates into a small error 
in $\alpha$.

\begin{figure}[t]
  \centering
  \vspace*{-0.2cm}
  \includegraphics[width=9cm]{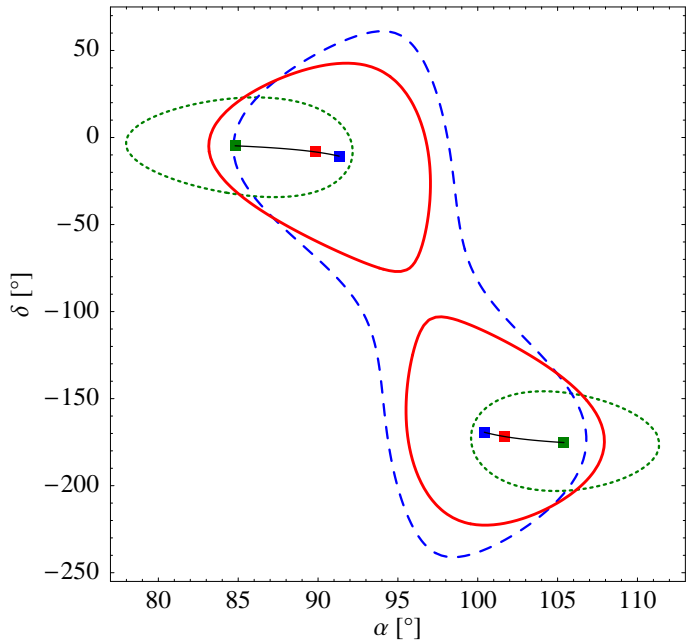}
  \caption{$\chi^2=1$ contours for $C_L,~S_L,~\R$ projected on the 
  $(\alpha,\delta)$ plane. Dotted, solid and dashed curves 
  correspond to $F=0.3,~0.9,~1.5$. The six points at focus 
  denote the solutions for vanishing errors in $C_L,~S_L,~\R$.  
  \label{fig:plot2}}
\end{figure}

In order to quantify the criterion excluding solution (2) for $\alpha$,
we study now the dependence of the two solutions for 
$\alpha$ near $\pi/2$ on the strong phase $\delta$. In Figure~\ref{fig:plot2} 
we plot the $\chi^2=1$ contours for $C_L,~S_L$ and $\R$
projected on the $(\alpha,\delta)$ plane. Three different values of 
$F$, $F=0.3,~0.9,~1.5$, are described by the dotted, solid and dashed 
curves, respectively. 
The upper and lower parts of the curves 
correspond to solutions (1) and (2) discussed above. 
The $\chi^2=1$ contours of the two solutions merge because of current 
experimental errors. The six points of focus for the three curves, marked 
on two almost 
parallel solid segments around 
$\delta=-8^\circ$ and $\delta = -172^\circ$, are obtained for
vanishing experimental errors in $C_L,~S_L$ and $\R$. The length of 
the segments gives the purely theoretical uncertainty in $\alpha$ 
originating in the range of  $F$. Figure~\ref{fig:plot2} shows that 
when including current experimental errors in $C_L,~S_L$ and $\R$ 
the two solutions (1) and (2) are presently distinguishable by 
$|\delta| < \pi/2$ and $|\delta| > \pi/2$, respectively. 
The additional requirement $|\delta| <\pi/2$, which excludes the second 
solution, will be relaxed considerably with more precise data on $C_L$, 
the error of which determines the uncertainty in $\delta$.

{\bf 6.} 
We conclude with a few comments about future improvements in the 
determination of $\alpha$. 
\begin{enumerate}
\item The theoretical error in the extracted value of $\alpha$
 depends weakly on the range assumed for the parameter $F$ and on
 the measurement of ${\cal R}$. 
 A resolution of the disagreement 
 between the BABAR and BELLE measurements of ${\cal R}$ will  
 be reassuring. More precise measurements of $S_L$ 
 will have direct impact on the experimental error $\pm 7^\circ$ 
 on $\alpha$, while more precise measurements of $C_L$ will eventually 
 reduce the phase assumption to a discrete choice.
 This may be compared to the isospin-method for extracting 
 $\alpha$ from $B\to\rho\rho$, where further reduction of the error 
 depends on what values the branching fractions will take.
 An intrinsic theoretical uncertainty at a level of a few degrees,
 caused in the isospin method by an $I=1$ final state originating from 
 the $\rho$ width~\cite{Falk:2003uq} and by $\rho-\omega$ 
 mixing~\cite{Gronau:2005pq}, may potentially be resolved by studying 
 with very large statistics the dependence of $B\to 4\pi$ decay distributions 
 on the invariant masses of pairs of pions near the $\rho$ mass.
\item Our suggestion for improving the determination of $\alpha$
 replaces the application of isospin bounds in $B\to\rho\rho$ by 
 theoretical input on the rough magnitude of $F$ 
 and a weak assumption about the the relative strong phase between 
 the penguin and tree amplitudes in $B^0\to\rho^+\rho^-$.
 Currently the assumption $|\delta| <\pi/2$ is required, but a weaker 
 condition will suffice in the future.   
 One possible test of this assumption consists of comparing 
 globally the pattern of tree-penguin interference in 
 $B\to \rho\rho$ and $B\to K^*\rho$ decays. 
\item Information about $\alpha$ is also obtained from 
 $B^0\to\rho^{\pm}\pi^{\mp}$ decays,
 which involve two ratios of penguin-to-tree amplitudes, 
 in $B^0\to\rho^+\pi^-$ and 
 $B^0\to\rho^-\pi^+$. SU(3) arguments relating these decays to $B\to
 K\rho$ and $B\to K^*\pi$~\cite{Gronau:2004tm}, and a calculation 
 based on QCD factorization~\cite{Beneke:2003zv} show that these 
 two ratios are small, in the range $0.1- 0.2$, being on the smaller 
 side in the second approach. The small ratios imply a 
 small deviation of $\alpha$ from the value of $\alpha_{\rm eff}$ 
 obtained in the absence of penguin amplitudes~\cite{Gronau:2004tm}. 
 Current data for time dependence in $B^0\to\rho^{\pm}\pi^{\mp}$, 
 given in terms of four observables, $C, S, \Delta C, 
 \Delta S$~\cite{Group:2006bi}, imply $\alpha_{\rm eff} = (94 \pm 4)^\circ$.  
 An SU(3)-derived bound on the effect of penguin amplitudes, 
 $|\alpha - \alpha_{\rm eff}|< 9^\circ$, implies $\alpha = 
 (94 \pm 10)^\circ$ when adding theoretical and 
 experimental errors in quadrature~\cite{Gronau:2004tm,Gronau:2004sj}. 
 A more precise determination, $\alpha = (94 \pm 7)^\circ$,
 follows from the observable $S$ alone using a QCD 
 factorization calculation for amplitudes and strong 
 phases~\cite{Beneke:2003zv}. Both determinations require 
 stronger assumptions than those made in this work. However, 
 the consistency of the most precise measurements of $\alpha$ 
 (hence, $\gamma$) is impressive, allowing us to conclude 
 that $\alpha$ is in the vicinity of $90^\circ$ within a 
 few degrees. 
\end{enumerate}

\vskip0.2cm\noindent
{\em Acknowledgements.} 
This work was supported in part
by the Israel Science Foundation under Grants No. 1052/04
and 378/05, and by the German--Israeli Foundation under Grant
No. I-781-55.14/2003. M.B. and J.R. would like to thank the 
Technion for its hospitality.

\subsubsection*{Note added in proof}

Shortly after this paper was submitted BABAR presented the new 
values $\b(K^{*0}\rho^+) =  10.0 \pm 1.7 \pm 2.4$ 
and $f_L(K^{*0}\rho^+) = 0.53 \pm 0.10 \pm 0.06$ \cite{Smith}, 
so that the BABAR and BELLE results are now in very good 
agreement [see (\ref{fLs})]. The new value of ${\cal R}$ 
equals  $0.0080 \pm 0.0023$ instead of (\ref{Rexp}). 
This leads to the following changes in 
our results: Since ${\cal R}$ and $F$ enter our analysis only in 
the combination ${\cal R}/F$ [see (\ref{calR})], the new value of 
${\cal R}$ and $F=0.9$ is equivalent to the old value of 
${\cal R}$ and $F=1.41$ in Figure~\ref{fig:plot}. It can be 
seen that within current experimental errors the two solutions 
corresponding to (1) and (2) in Table~\ref{tab:eightsolutions} 
overlap even for $F=0.9$, i.e. the corresponding $\chi^2=1$ 
contours in  Figure~\ref{fig:plot2} merge also for $F=0.9$. 
Separating the two solutions by requiring $\delta<\pi/2$, 
our final result (\ref{alpha}) now reads 
\[
\alpha = [91.2^{+9.1}_{-6.6}~({\rm exp})^{+1.2}_{-3.9}~({\rm th})]^\circ~.
\]
The larger experimental error is due to the fact that the two
solutions have merged. As discussed in the text, the forseeable improved 
measurement of $C_L$ will remedy this problem.

\end{document}